# Diffraction limited focusing and routing of gap plasmons by a metal-dielectric-metal lens

Brian S. Dennis[1], David A. Czaplewski[2], Michael I. Haftel[3], Daniel Lopez[2], Girsh Blumberg[1] and Vladimir Aksyuk[4,*]

[1]*Department of Physics and Astronomy, Rutgers, the State University of New Jersey, 136 Frelinghuysen Rd., Piscataway, NJ 08854, USA*
[2]*Argonne National Laboratory, Center for Nanoscale Materials, Argonne, Illinois 60439, USA*
[3]*Department of Physics, University of Colorado at Colorado Springs, Colorado Springs, Colorado 80918, USA*
[4]*Center for Nanoscale Science and Technology, National Institute of Standards and Technology, 100 Bureau Drive, Gaithersburg, MD 20899, USA*
[*]*vladimir.aksyuk@nist.gov*

**Abstract:** Passive optical elements can play key roles in photonic applications such as plasmonic integrated circuits. Here we experimentally demonstrate passive gap-plasmon focusing and routing in two-dimensions. This is accomplished using a high numerical-aperture metal-dielectric-metal lens incorporated into a planar-waveguide device. Fabrication via metal sputtering, oxide deposition, electron- and focused-ion- beam lithography, and argon ion-milling is reported on in detail. Diffraction-limited focusing is optically characterized by sampling out-coupled light with a microscope. The measured focal distance and full-width-half-maximum spot size agree well with the calculated lens performance. The surface plasmon polariton propagation length is measured by sampling light from multiple out-coupler slits.

**1. Introduction**

Two-dimensional (2D) optical demonstrations of propagating collective electronic oscillations localized to metal–dielectric interfaces, or surface plasmon polaritons (SPPs) [1], have been reported for more than a decade [2-4]. Passive in-plane focusing of SPPs was first demonstrated in 2005 using both dielectric optical elements [6] and circular and elliptical plasmonic structures

[7,8]. Since then other plasmonic systems have been used to demonstrate focusing including nanoparticle chains [9], holographic arrays [10-12], circular gratings [13], plasmo-fluidics [14], nano-corrals [15] and Luneberg and Eaton lenses [16].

Gap plasmons (GPs) exist when a metal–dielectric–metal (MDM) waveguide confines the electromagnetic energy transversely in the dielectric and metal layers [17]. Increasingly, demonstrations of plasmonic devices are being reported using GPs rather than SPPs and include dimple lenses [18], plasmon-mechanical couplers [19], opto-electronic [20,21] and nanomechanical phase-modulators [22], amplitude modulators [23], nanofocusers [24], lasers [25], resonators [26], absorbers [27,28], sharp bend waveguides [29], nanomechanical switches [30], and waveguide couplers [31]. In this work we combine both gap plasmons and passive surface plasmon focusing. We fabricate an MDM planar waveguide device with a 2D plano-convex lens to demonstrate GP focusing and manipulation of the imaging angle by under-filling different portions of the lens with a collimated GP beam. A comparison with optical theory is also discussed and good agreement with the focal length and the focused spot size is achieved. In contrast with GPs confined in vertical slot waveguides [20,21,32], GP modes in MDM waveguides are similar to the modes in planar dielectric photonics. Therefore it is possible to combine and transition between waveguided modes [30,33,34], extended 2D modes [18, 22], and free-space modes [31,35].

## 2. Device description

Figure 1a shows a schematic of the MDM focusing device installed in the experimental set up, which consists of a 780 nm laser that is fiber coupled to the top 10x objective of a modified inverted microscope. Out-coupled light is collected with the bottom objective and sent to a CCD camera.

The device is fabricated from an MDM Au/SiO$_2$/Au stack and has three distinct cross sections: 1) Au/SiO$_2$/Au; 2) Au/air/Au; and 3) air/Au. A free-space laser grating-couples to a GP mode in the slot waveguide. The GP is launched, in the air gap under the grating, propagates through an SiO$_2$ bridge-support, through the air gap under suspended bridges, through a Au/SiO$_2$/Au lens, enters an out-coupler region where the top Au and SiO$_2$ has been removed, and converts to a surface plasmon. Here there are five equidistant out-coupler slits in the bottom Au layer and the propagating SPP partially out-couples to light as it passes over the slits (Fig. 1b). The top view of the GP propagation path is shown in Fig. 1c, represented by the red dashed lines for the GP in the waveguide and solid red to show the SPP focused onto the second out-coupler slit. The SiO$_2$ portion of the lens is highlighted in pink for emphasis. The blue dashed lines represent the 2.5 μm undercut into the SiO$_2$, under openings milled into the top Au, that happens during the liquid acid etching. The suspended Au bridges (Figs. 1b and 1c) were previously reported to electro-mechanically modulate the GP phase [22]. In this application the slits that define the bridges are used only as liquid etching ports (discussed later) to create an air waveguide below the bridges.

## 3. Experimental

### 3.1 Nanofabrication

Figure 2 shows the process flow that was used to nanofabricate the device. Wafer pieces 20 mm x 20 mm were used, diced from a 500 μm thick borosilicate glass wafer. The dashed red line in Fig. 2a shows the cross section represented in Figs. 2b - 2f. The pre-lithographic stack is seen in Fig. 2b, with all metal deposition by room temperature sputtering and SiO$_2$ deposition by plasma-enhanced chemical vapor deposition at 180 °C. The stack deposited onto the glass wafer piece consists of a 10 nm Cr adhesion layer, a 220 nm Au layer, a 2 nm Ti adhesion layer, 220 nm of SiO$_2$, a 2 nm Ti adhesion layer and 220 nm of Au. A poly(methyl methacrylate)

(PMMA) positive e-beam resist layer was spun on top at ≈ 419 rad/s (4000 rpm) for 35 s and baked for 1 min at 180 °C for a layer thickness of ≈ 500 nm.

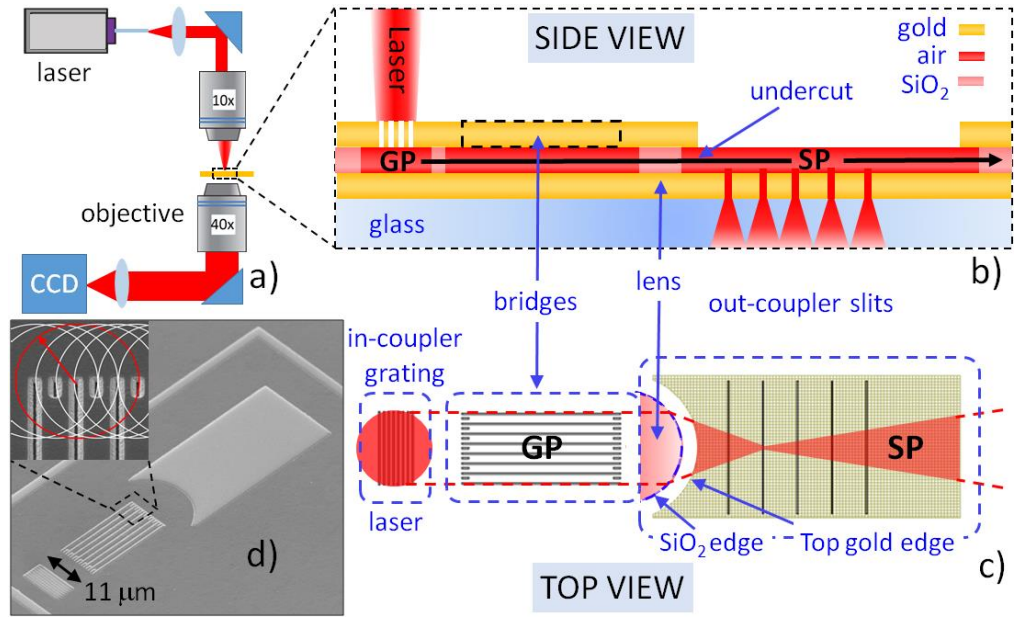

Fig. 1. Schematic of the experimental set up and MDM focusing device. a) A modified inverted microscope with a top excitation objective and a fiber coupled 780 nm laser. Laser light is focused from above, impinges on the device, and out-coupled light is collected from below and sent to a CCD camera. b) Side view: A focused free-space excitation laser grating couples to an MDM gap plasmon, propagates through an SiO$_2$ support structure, under 7 Au bridges, through the MDM lens, and converts to a *surface* plasmon at the bottom-Au/air interface before out-coupling to light at the five out-coupler slits. Light is collected from below through the glass substrate. The dashed black line represents the suspended bridges. c) Top View: The red dashed lines show the propagation path of the GP in the MDM waveguide, changing to solid red as the GP converts to a SPP after the lens where it is focused onto the 2nd slit. The blue dashed lines show the 2.5 μm undercut of the SiO$_2$ after release. Blue arrows point to the SiO$_2$ and top Au edges of the lens. d) Scanning electron micrograph of the device imaged at an angle. Seen from lower left to upper right are the in-coupler grating, bridges, lens, and out-coupler region (before the out-coupler slits were cut). The partial view of the square border is for electrical isolation and is not used in this application. The inset shows a close up of the bridge ends, with the overlapping 2.5 μm radius circles depicting the isotropic wet etch undercut boundaries from end points. The top edge of this etching pattern forms the plano- face of the lens and is mildly scalloped with features at the $\lambda_{GP}/15$ scale.

The lithographic pattern was written with 300 pA at 30 keV with area dose ≈ 18 μC/cm$^2$ for optimal exposure. Patterns were developed in 3:1 IPA:MIBK (isopropyl alcohol):(methyl isobutyl ketone) for 1 min, rinsed in a bath of IPA for 1 min and blown dry with compressed nitrogen (Fig. 2c). The pattern was transferred through the top Au and approximately half way into the SiO$_2$ layer with anisotropic argon ion milling, with the wafer rotating, the stage cooled to 10 °C and the incidence angle 10° off normal. The 20 cm diameter collimated ion beam had an acceleration voltage of 200 V to produce an ion current of 135 mA. Milling took place in six steps with one minute cooling in between. A thin PMMA layer ≈ 100 nm remained after ion milling (see Fig. 2d). Structures in the top Au layer were released by etching in a diluted liquid hydrofluoric acid bath (buffered oxide etch (BOE) 6:1) for 10 min, with subsequent rinsing in flowing water. The resultant horizontal undercut into the SiO$_2$ was ≈ 2.5 μm. IPA was added to the water (≈ 10 % by volume) to reduce the liquid surface tension to prevent suspended structures from drying and sticking to the bottom Au layer during transfer to a bath of IPA. Also to prevent the suspended top Au from sticking to the bottom Au, the device was dried in a $CO_2$ critical point dryer with the continuous-rinse exchange of IPA for liquid $CO_2$ taking ≈ 30 min (Fig. 2e). After release, ≈ 150 nm wide out-coupler slits were cut through the bottom Au layer and partially into the glass substrate with a Ga focused ion beam milling with 80 pA at 30 keV (Fig. 2f).

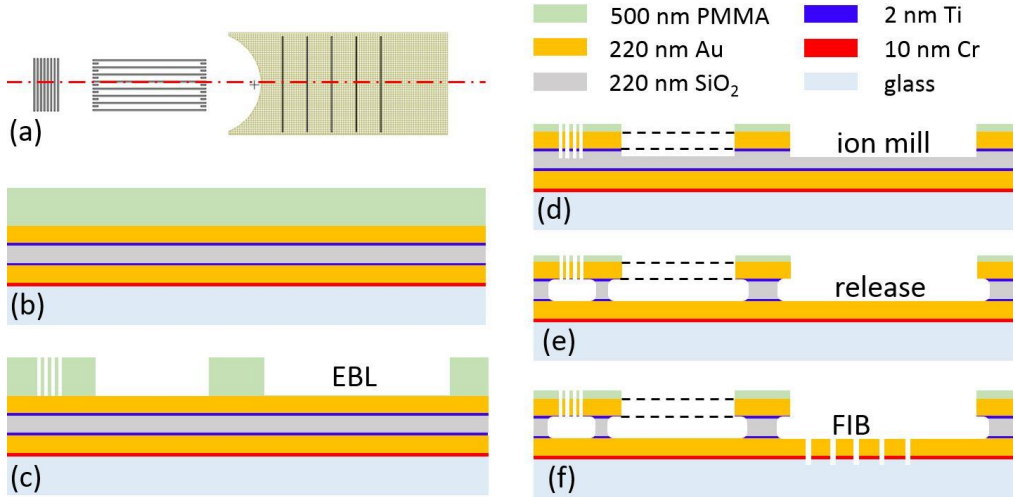

Fig. 2. Nanofabrication steps. (a) Top view schematic with a red dashed line between two central bridges showing the cross section depicted in side views (b) – (f). (b) MDM stack on glass substrate with PMMA on top. (c) Device structures are patterned into the PMMA positive resist with e-beam lithography. (d) Ar ion mill through the top Au and partially into the $SiO_2$. (e) Wet release in buffered oxide etch (BOE 6:1) for 10 min to remove the $SiO_2$ under the patterned areas followed by a $CO_2$ critical-point dry resulting in a 2.5 μm $SiO_2$ undercut. The horizontal dashed lines represent the suspended bridges. (f) Out-coupler slits are cut through the bottom Au and partially into the glass substrate with focused ion-beam milling.

## 3.2 Optical focusing demonstration

Figure 3 shows five optical images of the device taken from below, through the glass chip, while the GP is excited from above. Additional weak white light illumination from below (in reflection) makes the out-coupler slits visible as dark horizontal lines. The out-coupled light samples the focused SPP propagating over the five out-coupler slits. The images are overlaid on schematics of the device. The images experimentally demonstrate: 1) the ability of the MDM lens to transform a collimated GP into a focused SPP and; 2) that the focused SPP can be angularly directed via partial illumination of the lens' back aperture by changing the horizontal position of the excitation laser with respect to the in-coupler grating.

Fig. 3a shows that the excitation laser under fills the right hand side of the grating, couples to a collimated GP that propagates under the bridges, partially fills the right hand side of the MDM lens, and exits as a surface plasmon polariton into the out-coupler region, where the top Au and $SiO_2$ has been removed. The SPP is focused from the right hand side of the lens, onto the second slit, and propagates to the left, defocused. A schematic of the $SiO_2$ lens (blue) shows how the lens is under-filled by the GP. In Fig. 3b the laser is shifted slightly to the left on the grating, under-fills a more central part of the lens, and exits with less of an angle. Fig. 3c shows the laser incident on the center of the grating with the SPP centrally focused. A wide vertical pink arrow depicts the GP following the path outlined by the dashed vertical red lines. Figs. 3d and 3e show focusing with angles more to the right hand side as the laser is shifted to the left.

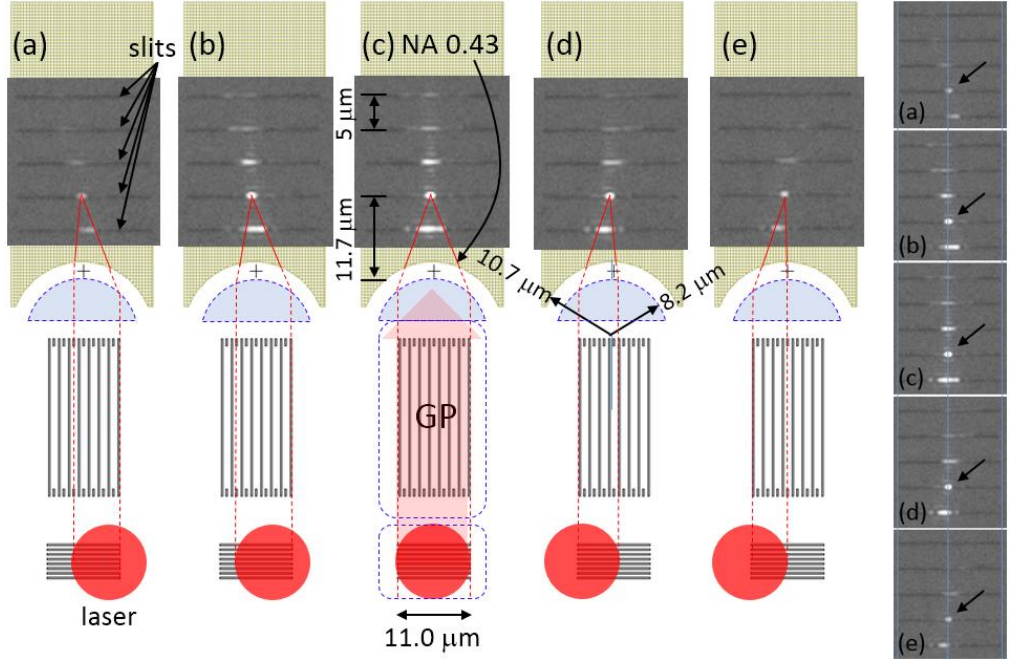

Fig. 3. Two-dimensional GP focusing. Left: Optical images (dark squares) of the out-coupled light sampled from five slits are superimposed over schematics of the device. The plasmon is focused on the second slit as it propagates over the slits, each separated by 5 μm. As the laser spot is shifted with respect to the in-coupler grating, the collimated GP shifts (vertical red dashed lines), under-filling different parts of the lens, resulting in tilted focused light. The laser is on the far right of the grating in (a), centered in (c), and to the far left in (e). Note that as the laser is shifted from center, the grating is under-filled, resulting in a narrower defocused GP and a less intense SPP at large angles. Shown in (c) are the distance from the vertex of the SiO$_2$ lens to the second slit (11.7 μm), the 0.43 NA plano-convex lens, and a schematic of the propagating GP. (d) The radii of the SiO$_2$ plano-convex and air meniscus lenses are shown. Red lines showing the solid angle of the focused SPP are guides for the eye. Far right: The black arrows in the vertical stack of images (a) - (e) point to the focused SPP on slit #2. Note that the focal point stays at the same point on the slit as the focus angle changes.

## 4. Analysis

### 4.1 Theoretical analysis ─ focusing

The two dimensional MDM lens (Fig. 1 top view) is plano-convex, with radii of curvature $r_1 = \infty$, $r_2 = -10.7$ μm ± 0.2 μm, vertical gap = 220 nm. The lens radii in this paper are derived from optical images using visible light. Their uncertainties are half of the full width at half maximum of the point spread function of the optical microscope with numerical aperture of 0.6 in the visible range. There is an undercut in the curved portion of the SiO$_2$, due to the release, that effectively transforms the lens into a compound lens composed of an Au/SiO$_2$/Au plano-convex lens and an Au/air/Au meniscus lens, each with a different index of refraction, separated by d = 0.

The Au/SiO$_2$/Au plano-convex lens has $r_1 = \infty$, $r_2 = -8.2 \pm 0.2$ μm. The index of refraction $n = 1.71$ can be obtained using the equations $k_{GP} = \beta = [k_2^2 + k_0^2 \varepsilon_2]^{1/2} = [k_0^2 \varepsilon_1 + 2(\varepsilon_1/\varepsilon_2 g)^2 - 2\varepsilon_1/\varepsilon_2 g[(\varepsilon_1/\varepsilon_2 g)^2 + k_0^2(\varepsilon_1 - \varepsilon_2)]^{1/2}]^{1/2}$ and $n = k_{GP}/k_0$, where $k_0 = 2\pi/\lambda_0$ is the laser wavevector in vacuum, $\varepsilon_1$ is the complex dielectric constant of the SiO$_2$, $\varepsilon_2$ is the dielectric constant of the Au and $g$ is the gap, the laser wavelength is $\lambda_0 = 780$ nm, $\varepsilon_1 = 1.5$, and $\varepsilon_2 = -22.4476 +$

1.36505$i$ (see ref. 22, Supplementary Sec.2 [22]). For a plano-convex lens, the lens formula simplifies to $1/f = (n − 1)[-1/r_2]$ and therefore $f_{plano} = 11.6$ µm.

The Au/air/Au meniscus lens has $r_1 = -8.2$ µm ± 0.2 µm, $r_2 = -10.7$ µm ± 0.2 µm, air gap = 220 nm, thickness t = 2.5 µm and index $n = 1.12$ can be calculated as above using $\varepsilon_1 = 1$. The meniscus lens will be weak due to the low index of refraction. The lens formula gives $1/f = (n - 1)[1/r_1 - 1/r_2 + (n - 1)t/nr_1r_2]$ and $f_{meniscus} = -328$ µm. Since t is approximately much less than $r_1$ and $r_2$, the combined focal length of the two lenses, separated by d = 0, is $1/f = 1/f_{plano} + 1/f_{meniscus}$ or $f_{calc} \approx 12.0$ µm. The distance from the vertex of the plano-convex SiO$_2$ lens to the second out-coupler slit is ≈ 11.7 µm (Fig. 3c). This means that the calculated focal distance is ≈ 0.3 µm further from the lens than the second slit location, but still agrees well with the calculated $f_{calc} \approx 12.0$ µm. Therefore the beam waist of the focused SPP may be narrower further from the slit.

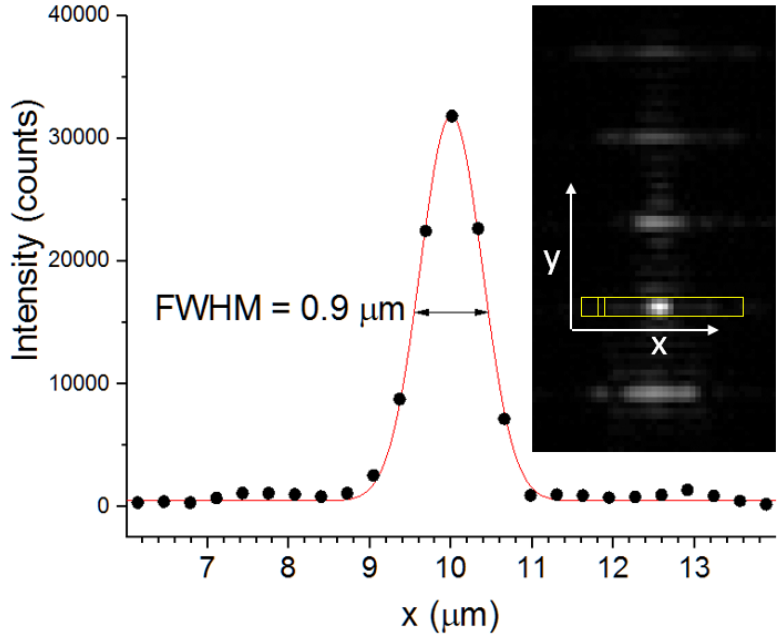

Fig. 4. Gaussian fit (red line) of the intensity profile of the focused SPP showing a diffraction limited spot wih 0.9 µm full width half max (FWHM). The profile is along the second slit in the x direction, taken from the yellow box in the inset. Each data point is a summation along y of three pixels. The statistical uncertainty is smaller than the symbol diameter. No white light was used so the slit is not seen. The contrast and brightness of the image were adjusted for clarity.

*4.2 Focused spot profiles*

The horizontal intensity profile of the out-coupled light, taken from the slit where the SPP was focused, was analyzed. The image used to obtain the profile was taken in the configuration of Fig. 3c, except there was no white light illumination. At each camera pixel in the x-direction the intensity of 3 pixels in the y-direction was vertically summed (Fig. 4 inset). A Gaussian fit was performed on the peak resulting in a full width half maximum (FWHM) of 0.90 µm ± 0.01 µm (Fig. 4) (unless otherwise noted, all experimental uncertainties reported are single standard deviation). The slits were illuminated in two different ways to measure the slit-widths used to calculate the objective point-spread-function. One configuration imaged out-coupled light coming through the slits. The other configuration used reflected white light illumination of dark slits. In each case the intensity profile in the direction normal to the slit was recorded and the measured FWHM was ≈ 0.5 µm, in agreement with a 40x objective with a numerical aperture of 0.6. Removing this broadening from the measurement, the plasmonic FWHM estimate is $(0.9^2 - 0.5^2)^{1/2}$ μm ≈ 0.75 μm. Conversion to this $1/e^2$ beam diameter gives a spot size of 1.28 μm.

The GP lens, with focal length $f_{calc}$ ≈ 12.0 μm, was illuminated with a D ≈ 11.0 μm wide (grating width) collimated GP for an effective NA = 0.43. The SPP emerging from the lens has wavelength $\lambda_{SPP}$ = 765 nm ($n_{air/Au}$ = 1.02). The diffraction limited $1/e^2$ beam diameter for a lens that is not overfilled by a Gaussian beam is $w_{calc} = 4\lambda_{SPP} f_{calc}/\pi D$ = 1.06 μm. Therefore the beam diameter measured at the second slit is about 20 % larger than $w_{calc}$. This is reasonable since the lens is slightly overfilled and the location of the second slit may be slightly out of focus of the compound plano-convex/meniscus lens. With such a large NA, the spherical lens' focusing performance may be improved with a slightly aspherical profile.

*4.3 Surface plasmon propagation length*

We report measurements of the SPP propagation length by microscopically sampling the light from multiple out-coupler slits [36] in a reference device without the lens. The reference is similar to the focusing device, but to make the transition from GP to SPP, the entrance to the out-coupler area is rectangular (Fig. 5), creating a narrow $SiO_2$ flat "plate" in place of the lens. In this case the SPP propagates as a collimated beam.

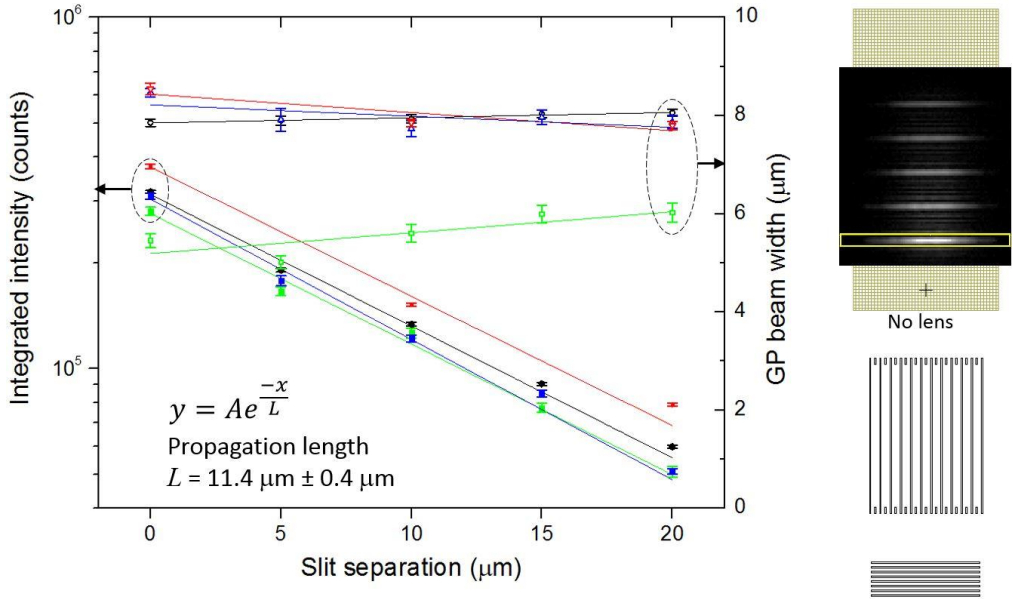

Fig. 5. Plots of the GP integrated intensity (left axis) and FWHM (right axis) from Gaussian fits of the out-coupled light from multiple-slit lensless devices. An example device schematic and superimposed image of out-coupled light is seen to the right of the plot. Intensity profiles were obtained from each slit, within the yellow rectangular box (example shown at the first slit), as described in Fig. 4. Left axis (log scale): Exponential fits (solid lines) of the data gives an average propagation length $L_{SPP}$ = 11.4 μm ± 0.4 μm. Note that the device depicted in red only has three out-coupler slits. Error bars are the standard deviations of the integrated area based on the Gaussian fits of the intensity profiles. Right axis: Linear fits of the data (solid lines) show how the GP beam widths change as they propagate and their level of collimation. Error bars are the standard deviations of the FWHM based on the Gaussian fits of the intensity profiles. Four devices were measured: 1) green: 7 bridges, 5 slits and an 11 μm wide grating; 2) red: 11 bridges, 3 slits and a 16 μm wide grating; 3) black: 11 bridges, 5 slits and a 16 μm wide grating; and 4) blue: 11 bridges, 5 slits and a 16 μm wide grating.

Intensity profiles from four devices, each with multiple slits, were obtained as previously described (Fig. 4) with Gaussian fits performed. Plotted in Fig. 5 are the integrated intensity and FWHM (GP beam width). Exponential decay constants, or propagation lengths, were extracted by fitting the intensity plots, giving individual propagation lengths and standard deviations for each device. From these four measurements the propagation length is determined to be $L_{SPP}$ = 11.4 μm ± 0.4 μm. The beam width data demonstrate the level of collimation as the GP propagates.

In an idealized case, where there is an absence of surface roughness, the SPP propagation length on a flat air/Au interface, with $\varepsilon_{air}$ = 1 and $\lambda_{SPP}$ = 765 nm, is given by $L_{SPP}$ = $1/2k''_x$ where $k''_x = [\omega\varepsilon_{Au}''/(2c(\varepsilon_{Au}'))^2]\cdot[\varepsilon_{Au}'/(\varepsilon_{Au}' + 1)]^{3/2}$ and $\varepsilon_{Au} = \varepsilon_{Au}' + i\cdot\varepsilon_{Au}''$ = -22.4476 + 1.36505$i$. This gives an ideal propagation length $L_{SPP}$ = 42 μm. The measured propagation length of the real device is smaller ($L_{SPP}$ = 11.4 μm), which we attribute to the effects of surface roughness from sputtered, processed Au. Scattering from individual slits is negligible within the experimental error, since a device with three out-coupler slits (1 slit every 10 μm) gives the same propagation length as those with five slits (1 slit every 5 μm) (Fig. 5).

*4.4 Scalloping*

Each bridge-end has an additional short medial slit (Fig. 1d) to increase the density of penetration points during the liquid etchant release. The purpose is to smooth out any scalloping effects, as the BOE will isotropically etch from every possible point. Etching forms the optical surface of both sides of the plano-convex $SiO_2$ lens at one end of the bridges and the support structure at the other end. Each bridge is 1.5 μm wide so the extra end-slits give a periodicity of etching penetration points of 0.75 μm. The etching circles from each point have a radius equal to the undercut length of 2.5 μm and overlap to form a very weak grating of $SiO_2$ cusps ≈ 30 nm in height with periodicity of d = 0.75 μm. The GP wavelength is 456 nm in the $SiO_2$, therefore the modulation depth is less than $\lambda_{GP}/15$. Since the ratio of the GP wavelength to the periodicity is almost unity, any 1st order diffraction from this weak grating into the $SiO_2$ would occur at $\theta = \sin^{-1}(\lambda_{GP}/d) = 37°$ and would not be seen as out-coupled light.

**4. Conclusion**

In conclusion, we have experimentally demonstrated that a planar lens can focus GPs to a Gaussian limited spot size and direct SPPs. This was accomplished by: 1) nanofabricating a GP device with a plano-convex MDM lens and a series of equally spaced out-couplers, and 2) collecting, imaging, and analyzing the out-coupled light. The imaging angle of the focused light was manipulated by changing the conditions of how a collimated GP under-fills the lens. The experimentally measured focal distance and beam waist agree with optical theory within experimental error. Using a lensless reference device, the SPP propagation length was measured by sampling out-coupled light from multiple slits. These results may contribute to the development of passive planar gap plasmonic devices.

**Acknowledgments**

This work has been supported by the Measurement Science and Engineering Research Grant Program of the National Institute of Standards and Technology (award nos. 70NANB14H259 and 70NANB14H030), the National Science Foundation DMR-1104884, and the Air Force Office of Scientific Research (grant no. FA9550-09-1-0698). Computational support from the Department of Defense High Performance Computation Modernization project is acknowledged. This work was performed, in part, at the Center for Nanoscale Materials, a US Department of Energy, Office of Science, Office of Basic Energy Sciences User Facility (contract no. DE-AC02-06CH11357).